\documentclass{iopart}
\usepackage[english]{babel}
\usepackage[utf8x]{inputenc}
\usepackage[T1]{fontenc}
\usepackage{amssymb,graphicx,url}

\newcommand{\LA}{\left\langle}
\newcommand{\RA}{\right\rangle}

\begin{document}
\title{Fast and accurate determination of modularity and its effect size}

\author{Santiago Treviño III$^{1,2}$, Amy Nyberg$^{1,2}$, Charo I.\ Del Genio$^{3,4,5,6}$ and Kevin E.\ Bassler$^{1,2,6}$}
\address{$1$ Department of Physics, 617 Science \& Research 1, University of Houston, Houston, Texas 77204-5005, USA}
\address{$2$ Texas Center for Superconductivity, 202 Houston Science Center, University of Houston, Houston, Texas 77204-5002, USA}
\address{$3$ Warwick Mathematics Institute, University of Warwick, Gibbet Hill Road, Coventry CV4 7AL, United Kingdom}
\address{$4$ Centre for Complexity Science, University of Warwick, Gibbet Hill Road, Coventry CV4 7AL, United Kingdom}
\address{$5$ Warwick Infectious Disease Epidemiology Research (WIDER) Centre, University of Warwick, Gibbet Hill Road, Coventry CV4 7AL, United Kingdom}
\address{$6$ Max Planck Institute for the Physics of Complex Systems, Nöthnitzer Str. 38, D-01187 Dresden, Germany}
\ead{bassler@uh.edu}
\begin{abstract}
We present a fast spectral algorithm for community detection
in complex networks. Our method searches for the partition with
the maximum value of the modularity via the interplay of several
refinement steps that include both agglomoration and division. 
We validate the accuracy of the algorithm
by applying it to several real-world benchmark networks. On
all these, our algorithm performs as well or better than any
other known polynomial scheme. This allows us to extensively
study the modularity distribution in ensembles of Erdős-Rényi
networks, producing theoretical predictions for means and variances
inclusive of finite-size corrections. Our work provides a way
to accurately estimate the effect size of modularity, 
providing a $z$-score measure of it
and enabling a more informative comparison of networks with different numbers
of nodes and links.
\end{abstract}
\maketitle

\section{Introduction}
Networked systems, in which the elements of a set of nodes
are linked in pairs if they share a common property, often
feature complex structures extending across several length
scales. At the lowest length scale, the number of links of
a node defines its degree $k$. At the immediately higher level,
the links amongst the neighbours of a node define the structure
of a local neighbourhood. The nodes in some local neighbourhoods
can be more densely linked amongst themselves than they are
with nodes belonging to other neighbourhoods. In this case,
we refer to these densely connected modules as communities.
A commonly used indicator of the prominence of community structure
in a complex network is its maximum modularity $Q$.
Given a partition of the nodes into modules, the modularity
measures the difference between its intra-community
connection density and that of a random graph null model~\cite{Alb02,New03,Boc06,Boc14}.
Highly modular structures have been found in systems of diverse
nature, including the World Wide Web, the Internet, social
networks, food webs, biological networks, sexual contacts
networks, and social network formation games~\cite{Pim79,Gar96,Fla02,Gir02,Eri03,Kra03,Lus04,Gui05,Del11,Tre12}.
In all these real-world systems, the communities
correspond to actual functional units.
For instance, communities in the WWW consist of
web pages with related topics, while communities
in metabolic networks relate to pathways and cycles~\cite{Fla02,Gui05,Pal05,Hus07}.
A modular structure can also influence the dynamical processes
supported by a network, affecting synchronization behaviour,
percolation properties and the spreading of epidemics~\cite{Res06,Are06,Del13}.
The development of methods to detect the community structure
of complex systems is thus a central topic to understand
the physics of complex networks~\cite{Dia07,Sch07,For10,Che14,Sob14}.

However, the use of modularity maximization to find communities
in networks presents some challenges and issues. The principal
challenge is that finding the network partition that maximizes
the modularity is an NP-hard computational problem~\cite{Bra08}.
Therefore, for a practical application, it is important to find
a fast algorithm that produces an accurate estimate of the maximum
modularity of any given network. Among the issues is that, in general, 
modularity itself does not allow for the
quantitative comparison of the modular structure between different networks.
For networks with the same number of nodes and links,
a higher modularity does indicate a more modular network structure.
However, this is not necessarily the case when
networks with different number of nodes or links are compared. 
In this paper, we present a spectral algorithm
for community detection based on modularity maximization and introduce
a method to estimate the effect size of modularity.
The algorithm we present incorporates both variations of the Kernighan-Lin
algorithm that remove constraints imposed on the resulting partition 
and an agglomeration step that can combine
communities. We validate the accuracy of our algorithm, which always
terminates in polynomial time, by applying it to a set of commonly
studied real-world example networks. We find that no other currently
known fast modularity maximizing algorithm performs better on any
network studied. We also use our algorithm to perform an extensive
numerical study of the distribution of modularity in ensembles of
Erdős-Rényi networks. Then, using our numerical results, we fit
finite-size corrections to theoretical predictions previously derived
for the mean of the distribution~\cite{Rei06,Rei06_2,Rei07} and to the
novel expression we derive for the variance, both of which are valid
in the large network limit. 
Finally, considering Erdős-Rényi networks as a null-model,
we obtain an analytic expression for a $z$-score
measure of the effect-size of modularity that is accurate for networks
of any size that have an average degree of~1 or more.
A quantitative comparison of $z$-scores
can be used to complement that of the modularities
of different networks, including those with different
numbers of nodes and/or links.

\section{Modularity and effect size}\label{modscore}
\begin{figure}
\centering
\includegraphics[width=0.25\textwidth]{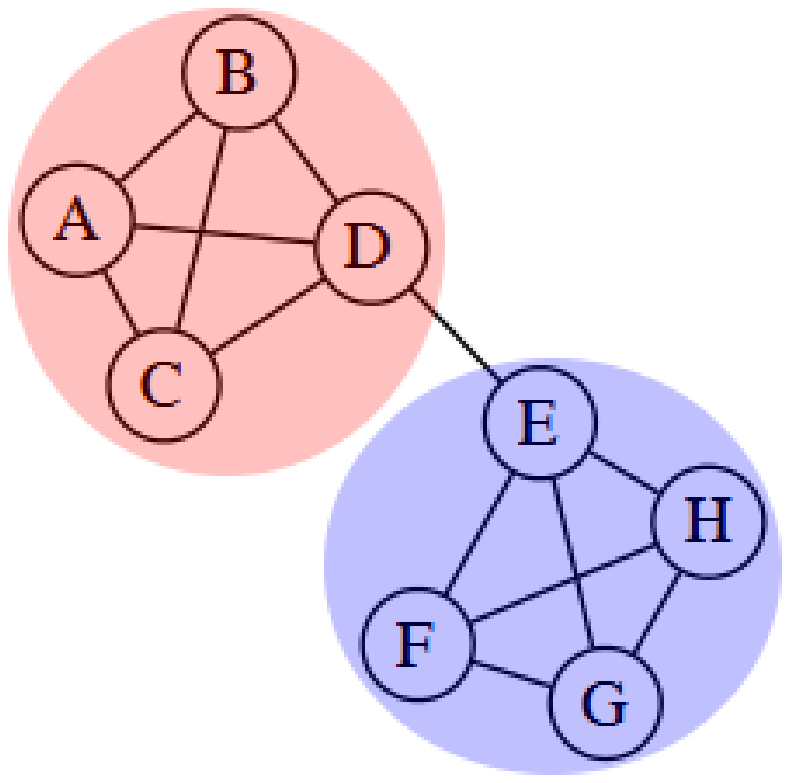}\quad\quad\quad
\includegraphics[width=0.4\textwidth]{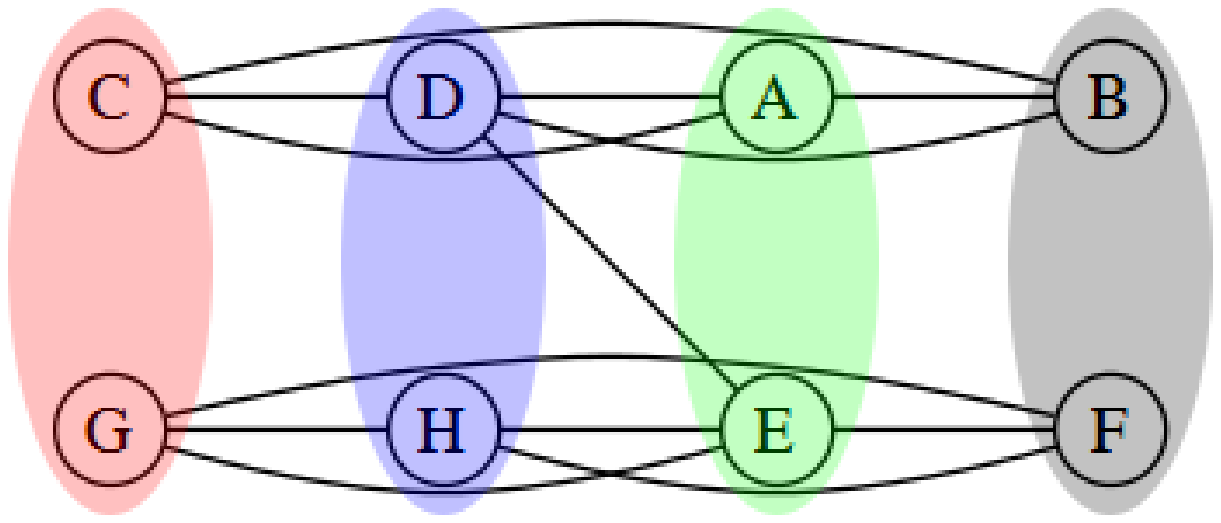}
\caption{\label{Fig1}Different partitions of the same network.
The partition on the left divides the network into two communities
with densely connected nodes. The partition on the right divides
the same network into modules consisting of disconnected nodes,
resulting in a low value of the modularity.}
\end{figure}

Given a network with $N$ nodes and $m$ links,
one can define a partition of the nodes by grouping
them into communities. Let $c_i$ indicate the
community to which node $i$ is assigned, and
let $\left\lbrace c\right\rbrace$ be the set
of communities into which the network was partitioned.
Then, the modularity $q$ of the partition is
\begin{equation}\label{moduldef}
q_{\left\lbrace c\right\rbrace} = \frac{1}{2m}{\sum}_{ij}\left(A_{ij} - \frac{k_ik_j}{2m}\right)\delta_{c_i,c_j}\:,
\end{equation}
where $k_i$ is the degree of node $i$, and $A$ is the adjacency
matrix, whose $\left(i,j\right)$ element is~1 if nodes $i$ and
$j$ are linked, and~0 otherwise. With this definition, the value
of $q$ is larger for partitions where the number of links within
communities is larger than what would be expected based on the
degrees of the nodes involved~\cite{New06}. Of course, even in
the case of a network with quite a well-defined community structure,
it is usually possible to define a partition with a small modularity.
For instance, one can artificially split the network into modules
consisting of pairs of unconnected nodes taken from different actual
communities (see Fig.~\ref{Fig1}). Thus, in order to properly characterize 
the community structure it is instead necessary
to find
the particular partition $\left\lbrace C\right\rbrace$ that maximizes
the modularity,
\begin{equation*}
 \left\lbrace C\right\rbrace = \arg\max_{\left\lbrace c\right\rbrace}\left\lbrace q_{\left\lbrace c\right\rbrace}\right\rbrace\:.
\end{equation*}
Henceforth, we indicate with $Q$ the maximum modularity of a network,
which is the modularity of the partition $\left\lbrace C\right\rbrace$:
\begin{equation*}
 Q = q_{\left\lbrace C\right\rbrace} = \max_{\left\lbrace c\right\rbrace}\left\lbrace q_{\left\lbrace c\right\rbrace}\right\rbrace = \max_{\left\lbrace c\right\rbrace}\left\lbrace\frac{1}{2m}{\sum}_{ij}\left(A_{ij} - \frac{k_ik_j}{2m}\right)\delta_{c_i,c_j}\right\rbrace\:.
\end{equation*}
The maximum modularity $Q$ corresponds to the particular
partition of the network that divides it into the most tightly
bound communities. However, simply finding this partition
is not sufficient to determine the statistical importance
of the community structure found.

To see this, consider an ensemble of random graphs
with a fixed number of nodes $N$ and a fixed number
of links $m$. As these networks are random, one can
safely say that they have no real communities. Then,
one could assume a vanishing average modularity $\LA q\RA_{\left\lbrace c\right\rbrace}$ on
the ensemble. However, the amount of community structure
is quantified by the extremal measure $Q$,
rather than $\LA q\RA_{\left\lbrace c\right\rbrace}$.
Thus, one cannot exclude \emph{a priori} the existence
of a partition with non-zero modularity
even on a completely random graph. This implies that
one can attach a fuller meaning to the maximum modularity
of a given network by comparing it to the expected
maximum modularity of an appropriate set of random
graphs. Then, the comparison defines an effect size
for the modularity, measuring the statistical significance
of a certain observed $Q$. Of course, the random graph
ensemble must be appropriately chosen to represent
a randomized version of the network analyzed.

A suitable random graph set for this study
is given by the Erdős-Rényi (ER) model ${\cal G}\left(N,p\right)$~\cite{Erd60}.
In the model, links between any pair of nodes
exist independently with fixed probability
$p$. As there is no other constraint imposed,
ER graphs are completely random, which makes
them a natural choice for a null model. Of
course, it is conceivable that another null
model could be used for specific types of
networks. In this case, one could generate
random ensembles of networks with a specified
set of constraints, using appropriate methods
such as degree-based graph construction~\cite{Del10,Kim12}.
To find the correct probability to use, we require
that the expected number of links in each
individual graph must equal the number of
links in the network we are studying. The
expected number of links in an Erdős-Rényi
network with $N$ nodes is
\begin{equation*}
 \LA m\RA = \frac{pN\left(N-1\right)}{2}\:.
\end{equation*}
Thus, the probability of connection must be
\begin{equation}\label{erp}
 p = \frac{2m}{N\left(N-1\right)}\:.
\end{equation}
Then, we can compare $Q$ with the expected
maximum modularity $\LA Q_{ER}\RA$ of the
ER ensemble thus defined. One simple way to
perform the comparison is calculating the
difference between $Q$ and $\LA Q_{ER}\RA$.
However, while this approach provides a certain
estimate of the importance of $Q$, it is not
entirely satisfactory. In fact, the same difference
acquires more or less significance depending
on the width of the distribution of $\LA Q_{ER}\RA$.
Then, it is a natural choice to normalize
the difference between maximum modularities
dividing it by the standard deviation $\sigma_{ER}$
of $\LA Q_{ER}\RA$
\begin{equation}\label{zscore}
 z = \frac{Q-\LA Q_{ER}\RA}{\sigma_{ER}}\:.
\end{equation}
The equation above defines a particular
measure of the effect size of $Q$ called
$z$-score. 
Positive $z$-scores indicate more modular structure than
expected in a random network, while
negative $z$-scores indicate less modular structure than
expected in a random network.

\section{Algorithm}\label{algo}
To find the maximum modularity partition
of a network, we introduce a variation of
the leading eigenvector method~\cite{New06,New06_2}.
The full algorithm provides a best guess
of the maximum modularity partition by progressively
refining the community structure. The general
idea is as follows. In the beginning, all
the nodes of the network are in the same
community. Then, one introduces the simplest
possible division, by splitting the network
into two different modules. The choice of
the nodes to assign to either module is
refined by several steps that are described
in detail below, and the whole
process is then repeated on each single
community until no improvement in modularity
can be obtained. 
A summary of the entire algorithm is
given in Subsection~\ref{summary}.

\subsection{Bisection}\label{bisection}
The first step in the algorithm
consists of the bisection of an
existing community. To find the
best bisection, we exploit the
spectral properties of the modularity
matrix $B$, whose elements are
defined by
\begin{equation*}
 B_{ij} = A_{ij} - \frac{k_ik_j}{2m}\:.
\end{equation*}
Substituting this into Eq.~\ref{moduldef},
we obtain an expression for the modularity
of a partition in terms of $B$:
\begin{equation}\label{modulwithB}
q_{\left\lbrace c\right\rbrace} = \frac{1}{2m}{\sum}_{ij}B_{ij}\delta_{c_i,c_j}\:.
\end{equation}
As we are considering splitting
a community into two, we can represent
any particular bisection choice
by means of a vector $s$ whose
element $s_i$ is $-1$ if node $i$
is assigned to the first community,
and $1$ if it is assigned to the
second. Then, using $\delta_{c_i,c_j} = \frac{1}{2}\left( s_i s_j + 1 \right)$,
Eq.~\ref{modulwithB} becomes
\begin{equation}\label{modulwithS}
q_{\left\lbrace c\right\rbrace} = \frac{1}{4m}{\sum}_{ij}B_{ij} s_i s_j\:.
\end{equation}
We can now express $s$ as a linear combination
of the normalized eigenvectors $v$ of $B$
\begin{equation*}
 s = \sum_{i=1}^N \alpha_i v_i\:,
\end{equation*}
so that Eq.~\ref{modulwithS} becomes
\begin{equation}\label{moduleigen}
q_{\left\lbrace c\right\rbrace} = \frac{1}{4m} \sum_{i=1}^N \alpha_i^2 \lambda_i\:,
\end{equation}
where $\lambda_i$ is the eigenvalue of $B$ corresponding
to the eigenvector $v_i$. From Eq.~\ref{moduleigen}, it
is clear that to maximize the modularity one could simply
choose $s$ to be parallel to the eigenvector $v_1$ corresponding
to the largest positive eigenvalue $\lambda_1$. However,
this approach is in general not possible, since the elements
of $s$ are constrained to be either~0 or~1. Then, the best
choice becomes to construct a vector $s$ that is as parallel
as possible to $v_1$. To do so, impose that the element
$s_i$ be~1 if ${v_1}_i$ is positive and~$-1$ if ${v_1}_i$
is negative.

This shows that the whole bisection
step consists effectively just of the
search for the eigenvalue $\lambda_1$
and its corresponding eigenvctor $v_1$.
As the modularity matrix $B$ is real
and symmetric, this is easily found.
For instance, one can use the well-known
power method, or any of other more
advanced techniques.
If $\lambda_1>0$, we build our best
partition as described above and compute
the change in modularity $\Delta Q$;
conversely, if $\lambda_1\leqslant 0$,
we leave the community as is.

The computational complexity
of the bisection step depends on the method
used to find $\lambda_1$. In the case of
the power method, it is ${\cal O}\left(Nm\right)$.

\subsection{Fine tuning}\label{KL}
Each bisection can often be improved using a variant of the Kernighan-Lin
partitioning algorithm~\cite{Ker70}. The algorithm considers moving each
node $k$ from the community to which it was assigned into the other, and
records the changes in modularity $\Delta Q'\left(k\right)$ that would result
from the move. Then, the move with the largest $\Delta Q'\left(k\right)$ is
accepted. The procedure is repeated $N$ times, each time excluding from consideration
the nodes that have been moved at a previous pass. Effectively, this tuning
step traverses a decisional tree in which each branching corresponds to one
node switching community. The particular path taken along the tree is determined
by choosing at each level the branch that maximizes $\Delta Q'\left(k\right)$.
Thus, it is clear that at each level $i$ in the tree the total modularity change
$\Delta Q'_i$ is simply the sum of the $\Delta Q'\left(k\right)$ considered
up to that point. When the process is over, and all the nodes have been eventually
moved, one finds the level $i^\ast$ with the maximum total modularity change:
\begin{equation*}
 i^\ast = \arg\max_i\left\lbrace\Delta Q'_i\right\rbrace\:.
\end{equation*}
If $\Delta Q'_{i^\ast}$ is positive, the partition is updated by switching
the community assignment for all the nodes corresponding to the branches
taken along the path from level~1 to~$i^\ast$; if, instead, $\Delta Q'_{i^\ast}$
is negative or zero, the original partition obtained after the previous step is left
unchanged. Finally, the entire procedure is repeated until it fails to produce
an increase in total modularity.

In~\ref{compkl}, we detail an efficient implementation
for this step, with a computational complexity of ${\cal O}(N^2)$
per update.

\subsection{Final tuning}\label{FT}
After the bisection of a module, the nodes are assigned
to two disjoint subsets. If the algorithm consisted solely
of these steps, the only possible refinements to the community
structure would be further division of these subsets. As
a result, the separation of two nodes into two different
communities would be permanent: once two nodes are separated,
they can never again be found together in the same community.
This kind of network partition has been proved to introduce
biases in the results~\cite{Sun09}. To avoid this, we introduce
in our algorithm a ``final tuning'' step, that extends
the local scope of the search performed by the fine tuning~\cite{Sun09,Che14,Sob14}.

To perform this step, we work on the network after all communities
have undergone the bisection and fine tuning steps individually. Then,
we consider moving each node from its current community to all other
existing communities, as well as moving it into a new community on
its own. For each potential move we compute the corresponding change
in modularity $\Delta Q''\left(k,\alpha\right)$, where $k$ is the node
being analyzed, and $\alpha$ is the target community. The particular
move yielding the largest $\Delta Q''$ is then accepted. The procedure
is repeated, each time not considering the nodes that have already
been moved, until all the nodes have been reassigned to a different
community. Similar to the fine tuning step, at the end of the process
one looks at the decisional tree traversed to find the intermediate
level $i^\ast$ with the largest total increase in modularity $\Delta Q''_{i^\ast}$. If
this is positive, the network partition is updated by permanently
accepting the node reassignments corresponding to the branches followed
up to level $i^\ast$. Conversely, if $\Delta Q''_{i^\ast}$ is negative
or zero, the starting network partition is retained. The whole procedure is
then repeated until it does not produce any further increase in modularity.

In~\ref{compft}, we detail an efficient implementation
for this step, with a computational complexity of ${\cal O}(N^3)$
per update.

\subsection{Agglomeration}\label{agg}
Both the fine tuning and the final tuning algorithms
refine the best guess for the maximum modularity partition
by performing local searches in modularity space. In
other words, both tuning algorithms only consider moving
individual nodes to improve the current partition. Here,
we introduce a tuning step that performs a global search
by considering moves involving entire communities. In
particular, the new step tries to improve a current
partition by merging pairs of communities.

A global search of this kind offers the possibility
of finding partitions that would be inaccessible to
a local approach. For example, assume that merging
two communities would result in an increase of modularity.
A local search could still be unable to find this new
partition because the individual node moves could force
it to go through partially merged states with lower modularity.
If this modularity penalty is large enough, the corresponding
moves will never be considered. Conversely, attempting
to move whole communities allows to jump over possible
modularity barriers in search for a better partition.

Thus, after each final tuning, we perform
an agglomeration step as follows. First,
we consider all existing communities and for each
pair of communities $\alpha$ and $\beta$ we compute
the change in modularity $\Delta Q'''\left(\alpha,\beta\right)$
that would result from their merger. Then, we
merge the two communities yielding the largest
$\Delta Q'''$. This process is repeated until only
one community is left containing all the nodes.
Then, we look at the decisional tree, as for the previous steps,
and find the level $i^\ast$ corresponding to the largest total
increase in modularity. If $\Delta Q'''_{i^\ast}$
is non-negative, we update the network partition
by performing all the community mergers resulting
in the intermediate configuration of the level
$i^\ast$. If, instead, $\Delta Q'''$ is negative,
the original partition is retained. Finally, the
whole procedure is repeated until no further improvement
can be obtained.

Note that in all the steps described we could encounter
situations where more than one move yields the same maximum
increase of modularity. In such cases we randomly extract
one of the equivalent moves and accept it. The only exception
is the determination of $i^\ast$ in the agglomeration step:
if multiple levels of the decisional tree yield the same
largest $\Delta Q'''$, rather than making a random choice,
we pick the one with the lowest number of communities.
This is intended to avoid spurious partitions of actual
communities as could result from the previous steps.

In~\ref{compaggl}, we detail an efficient implementation
for this step, with a computational complexity of ${\cal O}(N^3)$.

\subsection{Summary of the algorithm}\label{summary}
The steps described in the previous subsections
can be put in an algorithmic form, yielding our
complete method. Given a network of $N$ nodes:
\begin{enumerate}
 \item Initialize the community structure with all the nodes partitioned into a single community.
 \item Let $r$ be the current number of communities, and let $\alpha$ be a numerical label indicating
which community we are working on. Set $\alpha=1$.
 \item Attempt to bisect community $\alpha$ using the leading eigenvalue
method, described in Subsection~\ref{bisection}. Record the increase in modularity $\Delta Q$.
 \item Perform a fine-tuning step, as described in Subsection~\ref{KL}.
Record the increase in modularity $\Delta Q'$.
 \item If $\alpha<r$, increase $\alpha$ by~1 and go to step~(iii).
 \item Perform a final tuning step, as described in Subsection~\ref{FT}.
Record the increase in modularity $\Delta Q''$.
 \item Peform an agglomeration step as described in Subsection~\ref{agg}.
Record the increase in modularity $\Delta Q'''$.
 \item If the total increase in modularity $\Delta Q+\Delta Q'+\Delta Q''+\Delta Q'''$
is positive, repeat from step~(ii); otherwise, stop.
\end{enumerate}

Note that one is free to arbitrially set the tolerances
for each of the various numeric comparisons in the different steps of
the algorithm.
Every network will have a set of optimal tolerances,
which can be empirically determined, that will
yield the best results.
Generally, these tolerances should not be too low, as
they would make the algorithm behave like a hill
climbing algorithm. At the same time, they should not be too high,
as they would produce, effectively, a random search.

\subsection{Algorithm validation}
\begin{table}
\caption{\label{tab1}Algorithm accuracy validation.
The comparison between the maximum modularity found
by our algorithm ($Q$) and the best published result
($Q_{\mathrm{pub}}$) shows that no other modularity
maximizing scheme performs better than our method.
The benchmark networks used are, in order, the social
network in an American karate gym, the social network
of a community of dolphins in New Zealand, 
the network
of co-purchases of political books on Amazon.com in
2004, the word adjacency network in David Copperfield,
a collaboration network between jazz musicians, the
metabolic network in C. Elegans, the network of emails
exchanged between members of the Universitat Rovira~i~Virgili
in Tarragona, a network of trust in cryptographic key
signing, and a symmetrized snapshot of the structure
of the Internet at the level of autonomous systems,
as of July 22, 2006. The ``Time'' column
contains the time needed to complete a single run
of the algorithm on a stand-alone affordable workstation
at the time of writing.
The last column contains references to the methods
used to obtain the best result previously published 
for the corresponding network.}
\lineup
\begin{tabular}{@{}llllr@{\footnotesize{.}}lr@{\footnotesize{.}}lll}
\br
\footnotesize{Network}&\footnotesize{Nodes}&\footnotesize{Links}&\footnotesize{$Q$}&\multicolumn{2}{c}{\footnotesize{$z$-score}}&\multicolumn{2}{c}{\footnotesize{Time}}&\footnotesize{$Q_{\mathrm{pub}}$}&\footnotesize{Method}\\
\mr
\footnotesize{Karate~\cite{Zac77}}     &    \footnotesize{34} &    \footnotesize{78} & \footnotesize{$0.4198$} &    \footnotesize{1} & \footnotesize{68} & \footnotesize{0} & \footnotesize{45 ms} & \footnotesize{$0.4198$}& \footnotesize{\cite{Sob14,Duc05,Noa09,Goo10,Lem11}}\\
\footnotesize{Dolphins~\cite{Lus03}}   &    \footnotesize{62} &   \footnotesize{159} & \footnotesize{$0.5285$} &    \footnotesize{5} & \footnotesize{76} & \footnotesize{1} & \footnotesize{39 ms} & \footnotesize{$0.5276$} & \footnotesize{\cite{Goo10}}\\
\footnotesize{Books~\cite{KreXX}}      &   \footnotesize{105} &   \footnotesize{441} & \footnotesize{$0.5272$} &   \footnotesize{18} & \footnotesize{27} & \footnotesize{2} & \footnotesize{18 ms} & \footnotesize{$0.5272$} & \footnotesize{\cite{Sob14,Duc05,Goo10}}\\
\footnotesize{Words~\cite{New06}}      &   \footnotesize{112} &   \footnotesize{425} & \footnotesize{$0.3134$} &   \footnotesize{-3} & \footnotesize{51} & \footnotesize{5} & \footnotesize{65 ms} & \footnotesize{$0.3051$} & \footnotesize{\cite{Sob14}}\\
\footnotesize{Jazz~\cite{Gle03}}       &   \footnotesize{198} &  \footnotesize{2742} & \footnotesize{$0.4454$} &  \footnotesize{108} & \footnotesize{91} & \footnotesize{16} & \footnotesize{67 ms} & \footnotesize{$0.4454$} & \footnotesize{\cite{Sob14,Noa09}}\\
\footnotesize{C. Elegans~\cite{Duc05}} &   \footnotesize{453} &  \footnotesize{2025} & \footnotesize{$0.4526$} &   \footnotesize{21} & \footnotesize{97} & \footnotesize{155} & \footnotesize{85 ms} & \footnotesize{$0.4522$} & \footnotesize{\cite{Sob14,Sun09}}\\
\footnotesize{Emails~\cite{Gui03}}     &  \footnotesize{1133} &  \footnotesize{5045} & \footnotesize{$0.5827$} &   \footnotesize{70} & \footnotesize{89} & \footnotesize{1} & \footnotesize{30 s} & \footnotesize{$0.5825$} & \footnotesize{\cite{Sob14}}\\
\footnotesize{PGP~\cite{Bog04}}        & \footnotesize{10680} & \footnotesize{24316} & \footnotesize{$0.884$}  & \footnotesize{-144} & \footnotesize{17} & \footnotesize{43} & \footnotesize{33 min} & \footnotesize{$0.884$}  & \footnotesize{\cite{Noa09}}\\
\footnotesize{Internet\cite{NewUN}}    & \footnotesize{22963} & \footnotesize{48436} & \footnotesize{$0.6693$} & \footnotesize{-217} & \footnotesize{95} & \footnotesize{4} & \footnotesize{08 h} & \footnotesize{$0.6475$} & \footnotesize{\cite{Che14}}\\
\br
\end{tabular}
\end{table}
To validate the accuracy of our algorithm,
we applied it to a set of commonly studied
real-world unweighted undirected benchmark networks, and compared
its performance with the best result amongst
known methods. The results are shown in Table~1.
For each network we ran our algorithm a number of times,
and the best result we obtained is what is reported in the Table.
In each case, the best result was obtained 
within the first one hundred runs. For
the time estimates, we ran our algorithm on
a single core of an affordable, stand-alone
workstation with a single Intel${}^{®}$ Core${}^{™}$
i5-2400 CPU and 4~GB of RAM. The processor
is, at the time of writing, almost 4~years
old, having been introduced by the manufacturer
in January~2011.
In all cases considered, no other fast modularity
maximizing algorithm finds a more modular network
partition than the one identified by our method.
In fact, there are no reported results even from
simulated annealing, a slow algorithm, that exceed
ours. 

\section{Estimating the effect size}\label{scaling}
\begin{figure}
\centering
{\includegraphics[width=0.6\textwidth]{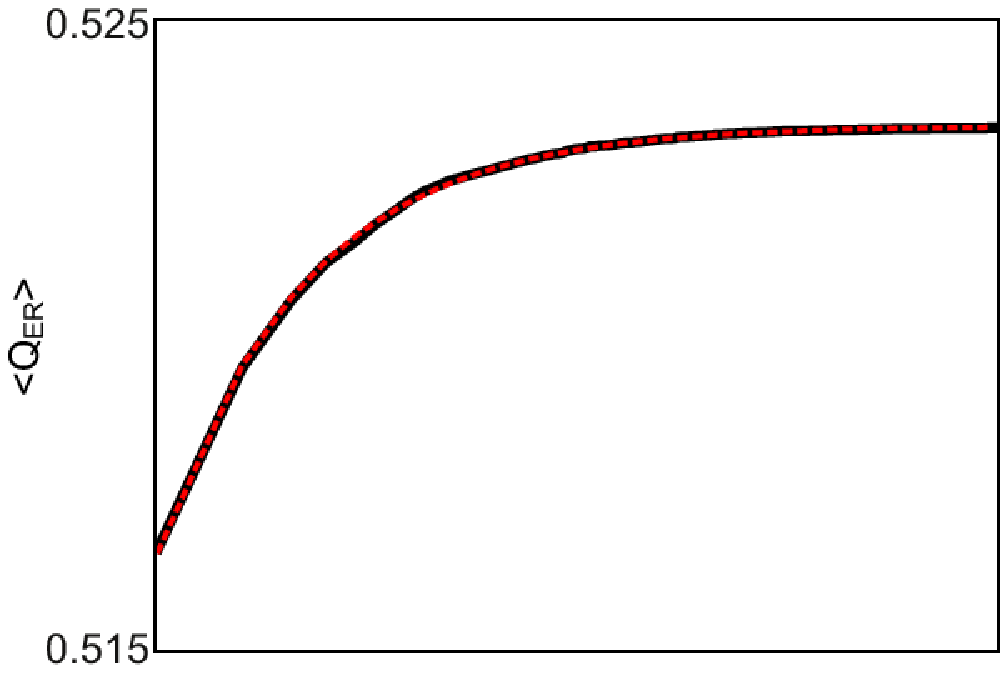}\vspace*{-20pt}}
\includegraphics[width=0.6\textwidth]{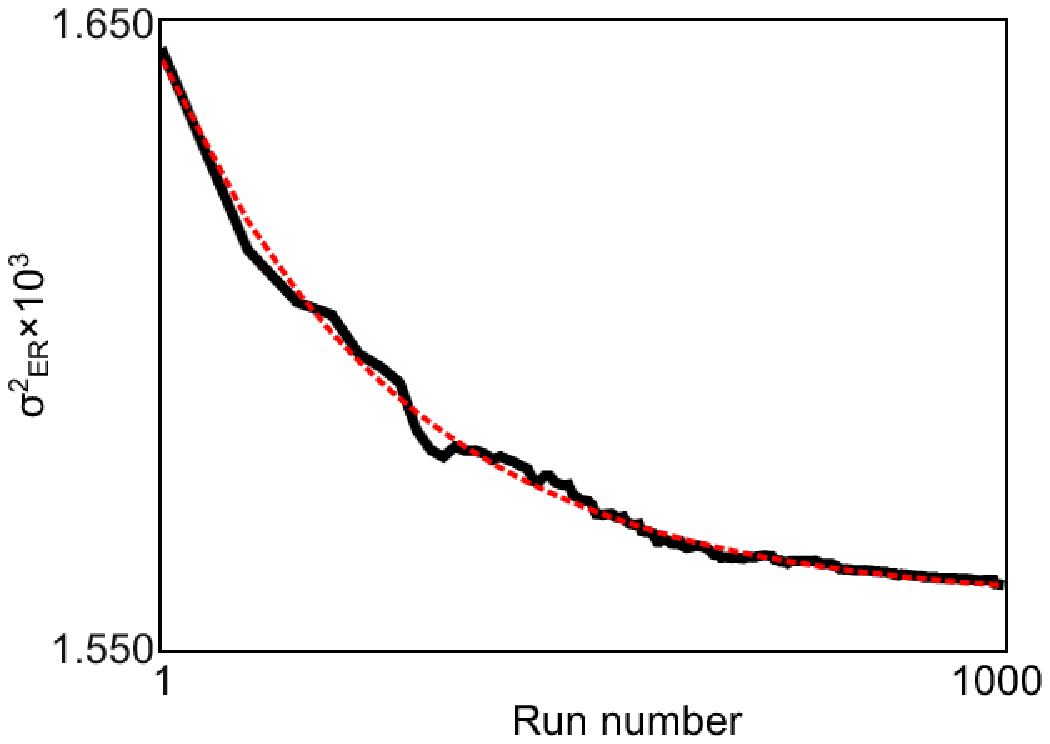}
\caption{\label{Fig2}Convergence of modularity distribution
for an ensemble of ER graphs with $N=50$ and $p=0.06$. The
top panel shows the best modularity measured after a given
number of runs; the bottom panel shows the respective variance.
The dashed red lines are power-law fits to the measured data.}
\end{figure}
As discussed in Section~\ref{modscore},
to compute the $z$-score for a given partition
of a network, we need to know the expected
maximum modularity $\LA Q_{ER}\RA$ of an
appropriately defined ER ensemble, and
its standard deviation $\sigma_{ER}$. To
find an expression for these quentities,
we start from the results in Refs.~\cite{Rei06,Rei06_2,Rei07},
which provide an estimate of $\LA Q_{ER}\RA$
for a generic ER ensemble ${\cal G}\left(N,p\right)$:
\begin{equation}\label{AVGQMAX}
\LA Q_{ER} \RA = 0.97 \sqrt{\frac{1-p}{Np}}\:.
\end{equation}
However, the equation above was derived under the assumptions
that $N \gg 1$ and $p\sim 1$. In other words, the estimate
is expected to be valid for large dense networks. Nevertheless,
in many real-world systems, networks are typically sparse~\cite{Del11_2},
and often their size is only few tens of nodes~\cite{Zac77,Lus03,Gle03}.
Therefore, to ensure the applicability of Eq.~\ref{AVGQMAX},
it is necessary to find appropriate scaling corrections. Finding
such corrections analytically is a very difficult problem. Thus,
here we employ a numerical approach.

First of all, to measure $\LA Q_{ER}\RA$ and $\sigma_{ER}$,
we performed extensive numerical simulations,
generating ensembles of Erdős-Rényi random graphs
with $N$ between~10 and~1000 and $p$ between~$1/N$
and~1. Then, we applied the algorithm described
in Section~\ref{algo} to each network in each
ensemble. However, as we discussed before,
the algorithm incorporates several elements
of randomness. In principle it can give
a different result every time it is run. Thus,
to estimate the expected maximum modularity
for each choice of $N$ and $p$, we ran the
algorithm 1000~times on each network, recording
after each run $r$ the largest value of modularity
obtained thus far, and computed ensemble averages
of $\LA Q_{ER}\RA\left(r\right)$ and $\sigma^2_{ER}\left(r\right)$.
The results show a fast convergence of the
quantities to their asymptotic value. To model
this convergence, we postulate that the difference
between the observed value and the asymptotic
one decays like a power-law with the number
of runs $r$:
\begin{eqnarray}
 \LA Q_{ER}\RA\left(r\right) &= \LA Q_{ER}\RA - Ar^{-B}\:,\nonumber\\
 \sigma^2_{ER}\left(r\right) &= \sigma^2_{ER} - Cr^{-D}\:.\nonumber
\end{eqnarray}
We can then fit the curves using $A$, $B$, $C$, $D$,
$\LA Q_{ER}\RA$ and $\sigma^2_{ER}$ as fit parameters,
as shown in Fig.~\ref{Fig2}, obtaining our estimate
for the asymptotic values.
In all cases studied, we find that the distribution of modularity
values is approximately Gaussian.

\begin{figure}
\centering
\includegraphics[width=0.6\textwidth]{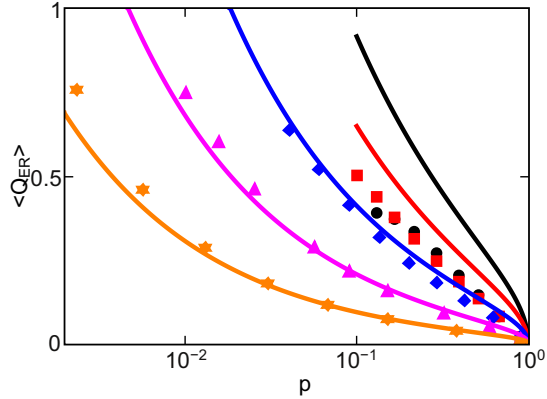}
\caption{\label{Fig3}Expected maximum modularity for ER ensembles.
The numerical data for $N=10$ (black dots), 20 (red squares), 50 (blue
diamonds), 200 (pink triangles) and 1000 (orange stars) show that the
predictions of Eq.~\ref{AVGQMAX} (like-coloured lines) are accurate
mostly for large dense networks.}
\end{figure}
Figure~\ref{Fig3} shows the final numerical results for $\LA Q_{ER}\RA$,
with the predictions of Eq.~\ref{AVGQMAX} for comparison.
For small system sizes, the measured
modularity is lower than that its theoretical prediction.
For larger systems, however, the approximation is effectively
in agreement with simulations, expect for lower values
of $p$, in the vicinity of the giant component transition.
This suggests the correction we need is twofold, consisting
of a multiplicative piece to scale down the prediction
for small systems, and an additive piece to account
for the case of sparse networks. Thus, an Ansatz for
the corrected form is
\begin{equation}
\LA Q_{ER}\RA = C_1 \cdot 0.97\sqrt{\frac{1-p}{Np}} + C_2\label{QC}\:.
\end{equation}

The simulation results seem to quickly approach the prediction
of Eq.~\ref{AVGQMAX} with increasing system size. Therefore, we
assume that $C_1$ is of the form
\begin{equation*}
C_1 = 1 - \lambda \mathrm e^{- \frac{N}{N_0}}\:.
\end{equation*}
Fitting these two parameters with the high-$p$ tail of the results yields
\begin{eqnarray}
\lambda &= \frac{7}{5}\:,\nonumber\\
N_0 &= 50\:\nonumber
\end{eqnarray}
Therefore, the multiplicative correction is 
\begin{equation}\label{C1}
C_1 = 1 - \frac{7}{5}\mathrm e^{-\frac{N}{50}}\:.
\end{equation}

\begin{figure}
\centering
\includegraphics[width=0.6\textwidth]{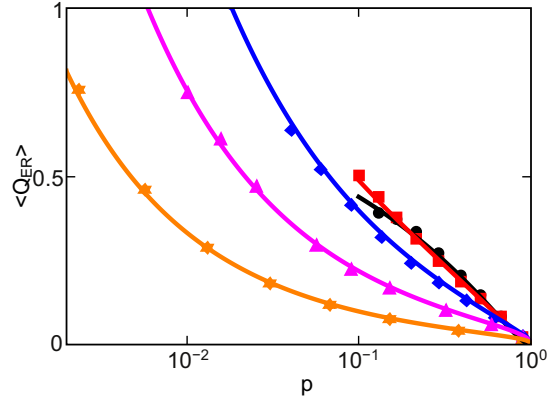}
\caption{\label{Fig4}Scaling corrections for expected maximum modularity.
The predictions of Eq.~\ref{MODCORFINAL} are accurate throughout the range
of $p$, and for all system sizes. The numerical data shown are for $N=10$
(black dots), 20 (red squares), 50 (blue diamonds), 200 (pink triangles)
and 1000 (orange stars).}
\end{figure}
The additive piece of the correction clearly depends
on $p$ and $N$. Thus, we start by assuming the general
form
\begin{equation}\label{C2}
C_2 = C_0 p^{\alpha}\left(1-p\right)^{\beta}N^\gamma\:,
\end{equation}
where the exponents $\alpha$, $\beta$, and $\gamma$ may depend on $N$.
Fitting these parameters yields
\begin{eqnarray}
C_0 &= 1\nonumber\\
\alpha &= -\frac{1}{6}\log\left(\frac{2}{5}N\right)\nonumber\\
\beta &= \frac{5}{4} \nonumber\\
\gamma &= -\frac{6}{5} + \frac{13}{15}\mathrm e^{-\frac{N}{100}}\:. \nonumber\\
\end{eqnarray}
To obtain the corrected expression for the expected
maximum modularity, substitute the parameter values
into Eq.~\ref{C2}, then substitute Eq.~\ref{C2} and
Eq.~\ref{C1} into Eq.~\ref{QC}:
\begin{equation}\label{MODCORFINAL}
\fl\LA Q_{ER}\RA = \left(1 - \frac{7}{5}\mathrm e^{-\frac{N}{50}}\right)0.97\sqrt{\frac{1-p}{Np}} + p^{-\frac{1}{6}\log{\left(\frac{2}{5}N\right)}}\left(1-p\right)^\frac{5}{4}N^{-\frac{6}{5}+\frac{13}{15}\mathrm e^{-\frac{N}{100}}}.
\end{equation}

\begin{figure}
\centering
\includegraphics[width=0.6\textwidth]{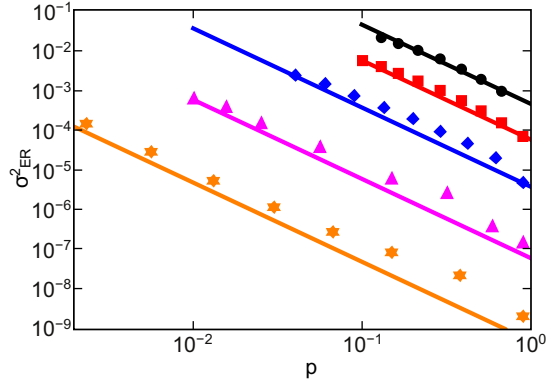}
\caption{\label{Fig5}Variance of the expected maximum modularity for ER ensembles.
The numerical data for $N=10$ (black dots), 20 (red squares), 50 (blue diamonds),
200 (pink triangles) and 1000 (orange stars) show that the predictions of Eq.~\ref{TEN}
(like-coloured lines) are accurate mostly for small sparse networks.}
\end{figure}
The predictions of Eq.~\ref{MODCORFINAL},
shown in Fig.~\ref{Fig4}, show a very good
agreement for all system sizes and all values
of $p$.
However, to compute the $z$-score of a given modularity measurement
on a particular network, we need to be able to express also the
variance of the modularity in the null model of choice. To
do so, we first use Eq.~\ref{AVGQMAX} to find the expected
form of the variance, using propagation of uncertainties.
Notice, however, that Eq.~\ref{AVGQMAX} was originally derived
in the framework of the ${\cal G}\left(N,m\right)$ ensemble,
in which the number of nodes $N$ and the number of edges $m$
are held fixed, rather than in the ${\cal G}\left(N,p\right)$
ensemble. Therefore, in finding an equation for the variance
of $\LA Q_{ER} \RA$, $p$ cannot be considered constant. Then,
\begin{equation}\label{SIX}
\sigma^2_{\LA Q_{ER} \RA} = \left(\partial_p \LA Q_{ER} \RA\right)^2\sigma^2_p\:.
\end{equation}
With $m$ fixed, one can write $p = \frac{2m}{N^2}$, hence
\begin{equation}\label{SEVEN}
\sigma_p^2 = \left(\partial_m p\right)^2\sigma^2_m = \frac{4}{N^4}\sigma^2_m\:.
\end{equation}
As $m$ is binomially distributed, its variance is
\begin{equation}\label{EIGHT}
\sigma^2_m = \frac{N^2}{2}p\left(1-p\right)\:.
\end{equation}
Substituting Eq.~\ref{EIGHT} into Eq.~\ref{SEVEN} yields
\begin{equation}\label{NINE}
\sigma_p^2 = \frac{2}{N}p\left(1-p\right)\:.
\end{equation}
Finally, substituting Eq.~\ref{NINE} into Eq.~\ref{SIX} one obtains
\begin{equation}\label{TEN}
\sigma^2_{\LA Q_{ER} \RA} = \frac{0.97^2}{2}\frac{1}{N^3p^2}\:.
\end{equation}

\begin{figure}
\centering
\includegraphics[width=0.6\textwidth]{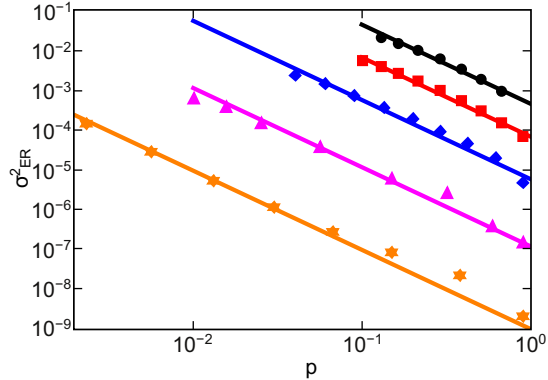}
\caption{\label{Fig6}Scaling corrections for variance of expected maximum modularity.
The predictions of Eq.~\ref{VARCOR} represent a substantial improvement for any $p$ and
all system sizes. The numerical data shown are for $N=10$ (black dots), 20 (red squares),
50 (blue diamonds), 200 (pink triangles) and 1000 (orange stars).}
\end{figure}

Once more, the results of the numerical simulations,
shown in Fig.~\ref{Fig5}, indicate that the actual variance
deviates from the theoretical prediction. Thus, also
in this case we need to find a correction. The deviation
of the measured variances from those predicted by means
of Eq.~\ref{TEN} rapidly increases with the size of
the network, apparently converging towards a constant.
Therefore, we postulate that the correction $C'$ to
Eq.~\ref{TEN} is multiplicative and has the form
\begin{equation*}
C' = C'_0 - \mathrm e^{-\varepsilon\left(N-N_0\right)}\:.
\end{equation*}
A fit of these parameters gives
$C'_0=2$, $\varepsilon=\frac{1}{50}$
and $N_0=10$. Thus, the final expression
for the variance of the expected
maximum modularity in a ${\cal G}\left(N,p\right)$
Erdős-Rényi ensemble is
\begin{equation}\label{VARCOR}
 \sigma^2_{\LA Q_{ER} \RA} = \left(2 - \mathrm e^{-\left(N-10\right)/50} \right)\frac{0.97^2}{2}\frac{1}{N^3p^2}\:.
\end{equation}
Again, the predictions of Eq.~\ref{VARCOR},
shown in Fig.~\ref{Fig6}, are in very good agreement
with the numerical simulations. We note, however,
that for values of $p$ greater than approximately
$0.15$ the numerically measured variance deviates slightly from the predicted
behaviour. 
In this region, we appear to slightly overestimate
the magnitude of the $z$-score. 
We postulate, however, that this is due to
an increased hardness in finding the best
partition for networks having this range of
connectivity.
Assuming this is case and our prediction of the
variance is correct in this region,
our estimate of the
magnitude of the $z$-score is accurate throughout the range.

With the corrections we developed, it is finally possible
to compute the $z$-score of a modularity measurement on any
particular network. First, one determines $p$ using Eq.~\ref{erp}.
Then, one uses Eqs.~\ref{MODCORFINAL} and~\ref{VARCOR}, to
calculate $\LA Q_{ER}\RA$ and $\sigma_{ER}$, respectively.
Finally, using these values, Eq.~\ref{zscore} yields the
$z$-score.

To further motivate the choice of the Erdős-Rényi
random graph ensemble as the natural null model for
network partitioning, we used the degree-based graph
sampling algorithm of Ref.~\cite{Del10} to construct
ensembles of networks with the same degree sequences
(SDS) as the benchmark systems we used for validation.
The comparison between the $z$-scores obtained with
the two approaches is shown in Table~\ref{tab2}. In
all cases, the SDS $z$-scores are more positive than
the ER ones. This strongly suggests that the SDS ensemble
underestimates the expected maximum modularity. The
reason for this behaviour is in the term $-\frac{k_ik_j}{2m}$
in Eq.~\ref{moduldef}, which estimates the number of
links between a node of degree $k_i$ and one of degree
$k_j$. This factor implicitly accounts for the possibility
of multiple edges in the networks, and therefore its
magnitude is larger than it should be. While this overestimate
is negligible for ER graphs, it becomes significant
for networks with degree distributions different from
those of random graphs. The effect is particularly
marked on scale-free networks, such as most of the ones we
analyzed here, since random networks with a power-law
degree distribution are known to be disassortative~\cite{Joh10,Wil14}.
These considerations suggest that the ER ensemble
is the correct null model to use for the calculation
of $z$-scores for modularity-based algorithms in most
community detection applications. 
\begin{table}
\caption{\label{tab2}Null model motivation.
Adopting as the null model the ensemble of
networks with the same degree sequence (SDS)
as the one studied always results in a more
positive $z$-score, indicating an underestimation
of the expected maximum modularity in the SDS
ensemble. The networks used here are the same
we used for the results shown in Table~\ref{tab1}.}
\lineup
\begin{indented}
\item[]\begin{tabular}{@{}lllr@{.}lr@{.}l}
\br
Network&Nodes&Links&\multicolumn{2}{c}{ER $z$-score}&\multicolumn{2}{c}{SDS $z$-score}\\
\mr
Karate~\cite{Zac77}     &    34 &    78 &    1 & 68 &   8 & 06\\
Dolphins~\cite{Lus03}   &    62 &   159 &    5 & 76 &  14 & 85\\
Books~\cite{KreXX}      &   105 &   441 &   18 & 27 &  37 & 70\\
Words~\cite{New06}      &   112 &   425 &   -3 & 51 &   2 & 07\\
Jazz~\cite{Gle03}       &   198 &  2742 &  108 & 91 & 150 & 20\\
C. Elegans~\cite{Duc05} &   453 &  2025 &   21 & 97 & 238 & 00\\
Emails~\cite{Gui03}     &  1133 &  5045 &   70 & 89 & 177 & 93\\
PGP~\cite{Bog04}        & 10680 & 24316 & -144 & 17 & 326 & 00\\
Internet\cite{NewUN}    & 22963 & 48436 & -217 & 95 & 358 & 52\\
\br
\end{tabular}
\end{indented}
\end{table}

\section{Conclusions}\label{concl}
In this paper we have presented practical methods for identifying 
community structure in complex networks and for quantifying whether
that structure is significant compared to what is expected in 
Erd\H{o}s-R\'{e}nyi networks. 
As such, our methods directly address the principal challenge and a major issue
with using modularity maximization to identify a network's 
community structure.
In particular, we have presented 
the best of any currently known algorithm 
for finding 
the network partition that maximizes modularity.
Then, making use of this algorithm and both existing and novel
analytical results, we found analytic expressions for the mean 
and standard deviation of the distribution
of modularity of ensembles of Erd\H{o}s-R\'{e}nyi networks.
Using these expressions, which apply
to all network sizes $N$ with average
connectivity $p>1/N$,
we have obtained an analytic transformation from modularity value
to a $z$-score that measures the effect size of modularity.

The conversion from modularity value to the $z$-score
of modularity effect size we have established is particularly 
noteworthy. Because of it, for the first time, one can
easily estimate the relative importance of the modular
structure in networks with different numbers of nodes or links.
This allows a new form of comparative network
analysis. For example, Table~1 lists the modularity $z$-scores
of the real-world test networks we used to validate our algorithm. 
Note that most of the networks have a $z$-score much greater
than 1, and thus their structure is substantially unlikely
to be due to a random fluctuation, with the
collaboration network of Jazz musicians
being by far the least random
of those studied. However,
the Key Signing network has a large {\em negative} $z$-score.
Thus, it is substantially less modular than a comparable
ER network. This indicates that, even though the network has
a very prominent modular structure, as evidenced by the large
modularity, its links are nonetheless much more evenly distributed
than expected if it were random. Similarly, we can say that
the word adjacency network in
``David Copperfield'' has a slightly less modular structure than
expected if random, and
the Karate Club network has a modular structure that could
still be attributed to a random fluctuation, although with
a probability of only about~5\%.
This form of analysis is clearly much more informative than one that
considers modularity alone. The difference is particularly striking,
for instance, with the Key Signing network, which has a very
high value of modularity, but a much less modular structure than a
comparable random network. The deeper level of insight the modularity
$z$-score provides
makes it ideal for the investigation of real-world networks, and
thus it will find broad application in the study of the Physics
of Complex Systems.

\ack
We would like to thank Florian Greil, Suresh Bhavnani
and Shyam Visweswaran for fruitful discussions.
ST, AN and KEB acknowledge funding from NSF
through Grant No.\ DMR-1206839, and by the AFOSR and DARPA
through Grant No.\ FA9550-12-1-0405. CIDG acknowledges support by
EINS, Network of Excellence in Internet Science,
via the European Commission's FP7 under Communications
Networks, Content and Technologies, grant No.~288021.

\appendix
\section{Computational complexity of the fine-tuning step}\label{compkl}
To estimate the worst-case computational complexity
of the fine-tuning step, we start by rewriting Eq.~\ref{modulwithS}
in vector form:
\begin{equation*}
q_{\left\lbrace c\right\rbrace} = \frac{1}{4m} s^{\mathrm T}\cdot B\cdot s\:.
\end{equation*}
In the following, to simplify the derivations,
sum is implied over repeated Roman (but not Greek)
indices. Then, it is
\begin{equation*}
q\equiv q_{\left\lbrace c\right\rbrace} = \frac{1}{4m} s_i B_{ij} s_j\:.
\end{equation*}

Now, consider switching the community assignment
of the $\alpha^{\mathrm{th}}$ node. This corresponds
to changing the sign of the $\alpha^{\mathrm{th}}$
component of $s$: $s_\alpha\rightarrow -s_\alpha$.
Thus, the new state vector is
\begin{equation*}
s' = s + \Delta s = s + \left(\begin{array}{c}
0 \\
\vdots \\
0 \\
-2s_\alpha \\
0 \\
\vdots\\
0
\end{array}\right)\:.
\end{equation*}
Then, the new value of the modularity is
\begin{eqnarray*}
q'\equiv q_{\left\lbrace c'\right\rbrace} & = \frac{1}{4m} s'^{\mathrm T}\cdot B\cdot s'\\
& = \frac{1}{4m}\left(s^{\mathrm T}\cdot B\cdot s + {\Delta s}^{\mathrm T}\cdot B\cdot s + s^{\mathrm T}\cdot B\cdot\Delta s + {\Delta s}^{\mathrm T}\cdot B\cdot\Delta s\right)\\
& = q + \frac{1}{4m} \left(-2s_\alpha B_{\alpha j}s_j-2s_iB_{i\alpha}s_\alpha+4s_\alpha B_{\alpha \alpha}s_\alpha\right)\\
& = q - \frac{1}{m} s_\alpha B_{\alpha i}s_i + \frac{1}{m} B_{\alpha \alpha}\:,
\end{eqnarray*}
where we have used the fact that $B$ is symmetric and $s_\alpha^2=1$.

Next, define the vector $W$ as $W\equiv B\cdot s$, so that its components
are
\begin{equation*}
W_i = B_{ij}s_j\:.
\end{equation*}
Then, we have
\begin{equation*}
q' = q - \frac{1}{m} s_\alpha W_\alpha + \frac{1}{m} B_{\alpha \alpha}\:.
\end{equation*}

Now, consider making a second change in a component
of $s$, say the $\beta^{\mathrm{th}}$ component, with
$\beta \neq \alpha$. The change is $s_\beta \rightarrow -s_\beta$.
Thus, the new state vector is
\begin{equation*}
s'' = s + \Delta s' = s + \left(\begin{array}{c}
0 \\
\vdots \\
0 \\
-2s_\alpha \\
0 \\
\vdots\\
0 \\
-2s_\beta \\
0 \\
\vdots\\
0
\end{array}\right)\:.
\end{equation*}
Then, the new value of modularity is
\begin{eqnarray*}
q''\equiv q_{\left\lbrace c''\right\rbrace} & = \frac{1}{4m} s''^{\mathrm T}\cdot{\mathbf B}\cdot s''\\
& = \frac{1}{4m}\left( s^{\mathrm T}\cdot B\cdot s + \Delta s'^{\mathrm T}\cdot B\cdot s + s^{\mathrm T}\cdot B\cdot\Delta s' + \Delta s'^{\mathrm T}\cdot B \cdot\Delta s'\right)\\
& = q + \frac{1}{4m}(-2s_\alpha B_{\alpha j}s_j-2s_\beta B_{\beta j}s_j-2s_iB_{i\alpha}s_\alpha-2s_iB_{i\beta}s_\beta\\
& \quad +4s_\alpha B_{\alpha \alpha}s_\alpha+4s_\beta B_{\beta \alpha}s_\alpha+4s_\alpha B_{\alpha \beta}s_\beta+4s_\beta B_{\beta \beta}s_\beta)\\
& = q - \frac{1}{m} s_\alpha B_{\alpha i}s_i - \frac{1}{m} s_\beta B_{\beta i}s_i + \frac{1}{m} B_{\alpha \alpha} + \frac{1}{m} B_{\beta \beta} + \frac{2}{m}s_\beta B_{\beta \alpha} s_\alpha\\
& = q' - \frac{1}{m} s_\beta B_{\beta i}s_i + \frac{1}{m} B_{\beta \beta} + \frac{2}{m} s_\beta B_{\beta \alpha} s_\alpha\\
& = q' - \frac{1}{m} s_\beta W_\beta + \frac{1}{m} B_{\beta \beta} + \frac{2}{m} s_\beta B_{\beta \alpha} s_\alpha\\
& = q' - \frac{1}{m} s_\beta W'_\beta + \frac{1}{m} B_{\beta \beta}
\end{eqnarray*}
where
\begin{equation*}
W'_\beta = W_\beta-2 B_{\beta \alpha} s_\alpha\:.
\end{equation*}

Generalizing to the $(n+1)^{\mathrm{th}}$ change,
\begin{equation*}
q^{(n+1)} = q^{(n)} - \frac{1}{m} s_{\alpha^{(n+1)}}W^{(n)}_{\alpha^{(n+1)}} + \frac{1}{m} B_{\alpha^{(n+1)} \alpha^{(n+1)}}\:,
\end{equation*}
where
\begin{equation*}
W^{(n)}_{\alpha^{(n+1)}} = W^{(n-1)}_{\alpha^{(n+1)}} - 2\sum_{p=1}^{n}B_{\alpha^{(n+1)} \alpha^{(p)}} s_{\alpha^{(p)}}\:.
\end{equation*}

Note that we need to calculate $W^{(n)}_{\alpha^{(n+1)}}$
for all possible remaining unchanged $\alpha^{(n+1)}$. Rewrite
this as
\begin{equation*}
W^{(n)}_{\alpha^{(n+1)}} = W^{(n-1)}_{\alpha^{(n+1)}} - \Delta W^{(n)}_{\alpha^{(n+1)}}\:,
\end{equation*}
where
\begin{equation*}
\Delta W^{(n)}_{\alpha^{(n+1)}} = 2 \sum_{p=1}^{n} B_{\alpha^{(n+1)} \alpha^{(p)}} s_{\alpha^{(p)}}\:.
\end{equation*}
But then
\begin{equation*}
\fl\Delta W^{(n)}_{\alpha^{(n+1)}} = 2 B_{\alpha^{(n+1)} \alpha^{(n)}} s_{\alpha^{(n)}} + 2 \sum_{p=1}^{n-1} B_{\alpha^{(n+1)} \alpha^{(p)}} s_{\alpha^{(p)}} = 2B_{\alpha^{(n+1)} \alpha^{(n)}} s_{\alpha^{(n)}} + \Delta W^{(n-1)}_{\alpha^{(n+1)}}
\end{equation*}

So, the fine-tuning algorithm can be implemented as follows: (prior knowledge of $q$ and $s$ is assumed)
\begin{enumerate}
\item Calculate $W^{(0)}_i = B_{ij}s_j$ for all $i$.
\item Calculate $q^{(1)} = q + \frac{1}{m} \left( B_{\beta \beta} - s_\beta W^{(0)}_\beta \right)$
for all $\beta$,  and choose the one that results in the largest value of $q^{(1)}-q$.
Define that value of $\beta$ to be $\alpha^{(1)}$.
\item Define $\Delta W^{(0)}_i = 0$ for all $i$.
\item Set $n=1$.
\item Calculate $\Delta W^{(n)}_i = \Delta W^{(n-1)}_i + 2B_{i\alpha^{(n)}}s_{\alpha^{(n)}}$
for all $i$ except $\left\lbrace \alpha^{(1)}, \ldots, \alpha^{(n)}\right\rbrace$.
\item Calculate $W^{(n)}_i = W^{(n-1)}_i - \Delta W^{(n)}_i$
for all $i$ except $\left\lbrace \alpha^{(1)}, \ldots, \alpha^{(n)}\right\rbrace$.
\item Calculate $q^{(n+1)} = q^{(n)} + \frac{1}{m} \left( B_{\beta \beta} - s_\beta W^{(n)}_\beta \right)$
for all $\beta$ except $\left\lbrace\alpha^{(1)}, \ldots, \alpha^{(n)}\right\rbrace$,
and choose the one that results in the largest value of $q^{(n+1)}-q^{n}$. Define that
value of $\beta$ to be $\alpha^{(n+1)}$.
\item If $n+1 < N$, set $n=n+1$ and go to step~(v).
\end{enumerate}

To estimate the computational complexity of the fine-tuning algorithm,
consider the complexity of each step:
\begin{itemize}
\item Using sparse matrix methods, Step~(i) is ${\cal O}(m)$.
Thus, in the worst case, its complexity is ${\cal O}(N^2)$.
\item Steps~(ii) and~(iii) are both ${\cal O}(N)$.
\item Step~(iv) is ${\cal O}(1)$.
\item Steps~(v) through~(viii) are ${\cal O}(N)$,
but are repeated ${\cal O}(N)$ times.
\end{itemize}
Thus, the total worst case complexity of one fine-tuning update is ${\cal O}(N^2)$.

Note that when applying the above treatment
to the bisection of a particular module of
a network, one should not disregard links involving
nodes that do not belong to the module considered.
Thus, the degrees of the nodes involved in
the calculation should not be changed and should
account for all the links incident to them~\cite{New06_2}.

\section{Computational complexity of the final-tuning step}\label{compft}
Consider a nonoverlapping partitioning of $N$ nodes
into $r$ communities. Then represent the partitioning
as an $N \times r$ matrix S where
\begin{equation*}
S_{ij} = \left\lbrace
\begin{array}{cl}
1 & \mbox{if node $i$ is in community $j$} \\
0 & \mbox{otherwise}\:.
\end{array}
\right.
\end{equation*}

Then, the modularity is
\begin{equation*}
q = \frac{1}{2m}\Tr\left(S^{\mathrm T}\cdot B\cdot S\right) = \frac{1}{2m}S_{ki}^{\mathrm T} B_{ij} S_{jk} = \frac{1}{2m}S_{ik} B_{ij} S_{jk}\:,
\end{equation*}
again we are implying sum over repeated Roman (but not Greek) indices.
Please also note our use of notation in what follows. Indices with $\alpha$
and $a$ designate one of the $N$ nodes. Indices with $\beta$ and $b$ designate
one of the $r$ communities. Thus, $S_{\beta \alpha}$ and $S_{\beta i}$
are elements of an  $r \times N$ matrix, while $S_{\alpha \beta}$ and
$S_{i \beta}$ are elements of an  $N \times r$ matrix. Also, by $1_{\alpha \beta}$
we indicate an matrix element $1$ in position $(\alpha,\beta)$. Note that
$1_{\beta \alpha}$ and $1_{\beta i}$ are unit valued elements of an $r \times N$
matrix, while $1_{\alpha \beta}$ and $1_{i \beta}$ are unit valued elements
of an $N \times r$ matrix.

Now, consider making a change in the community assignment
of one node, say $\alpha$, from community $\beta$ to $\beta'$,
with $\beta' \neq \beta$. Then, the new state matrix is
\begin{equation*}
\fl S' = S + \Delta S = S +\left(
\begin{array}{ccccccccccc}
0      & \cdots & 0      & 0                 & 0      & \cdots & 0      & 0                 & 0      & \cdots & 0\\
\vdots & \ddots & \vdots & \vdots            & \vdots & \ddots & \vdots & \vdots            & \vdots & \ddots & \vdots\\
0      & \cdots & 0      & -1_{\alpha \beta} & 0      & \cdots & 0      & 1_{\alpha \beta'} & 0      & \cdots & 0\\
0      & \cdots & 0      & 0                 & 0      & \cdots & 0      & 0                 & 0      & \cdots & 0\\
\vdots & \ddots & \vdots & \vdots            & \vdots & \ddots & \vdots & \vdots            & \vdots & \ddots & \vdots\\
0      & \cdots & 0      & 0                 & 0      & \cdots & 0      & 0                 & 0      & \cdots & 0\\
\end{array}
\right)\:.
\end{equation*}
Equivalently, we can write
\begin{eqnarray*}
\fl S' = S - \Delta S_- + \Delta S_+ = S-\left(
\begin{array}{ccccccccccc}
0      & \cdots & 0      & 0                 & 0      & \cdots & 0      & 0                 & 0      & \cdots & 0\\
\vdots & \ddots & \vdots & \vdots            & \vdots & \ddots & \vdots & \vdots            & \vdots & \ddots & \vdots\\
0      & \cdots & 0      & 1_{\alpha \beta}  & 0      & \cdots & 0      & 0                 & 0      & \cdots & 0\\
0      & \cdots & 0      & 0                 & 0      & \cdots & 0      & 0                 & 0      & \cdots & 0\\
\vdots & \ddots & \vdots & \vdots            & \vdots & \ddots & \vdots & \vdots            & \vdots & \ddots & \vdots\\
0      & \cdots & 0      & 0                 & 0      & \cdots & 0      & 0                 & 0      & \cdots & 0\\
\end{array}
\right)\\
+ \left(
\begin{array}{ccccccccccc}
0      & \cdots & 0      & 0                 & 0      & \cdots & 0      & 0                 & 0      & \cdots & 0\\
\vdots & \ddots & \vdots & \vdots            & \vdots & \ddots & \vdots & \vdots            & \vdots & \ddots & \vdots\\
0      & \cdots & 0      & 0                 & 0      & \cdots & 0      & 1_{\alpha \beta'} & 0      & \cdots & 0\\
0      & \cdots & 0      & 0                 & 0      & \cdots & 0      & 0                 & 0      & \cdots & 0\\
\vdots & \ddots & \vdots & \vdots            & \vdots & \ddots & \vdots & \vdots            & \vdots & \ddots & \vdots\\
0      & \cdots & 0      & 0                 & 0      & \cdots & 0      & 0                 & 0      & \cdots & 0\\
\end{array}
\right)\:.
\end{eqnarray*}

Thus, using the same convention as in the previous appendix, the new value of the modularity is
\begin{eqnarray*}
q' & = \frac{1}{2m}\Tr\left( S'^{\mathrm T}\cdot B\cdot S'\right)\\
& = \frac{1}{2m}\Tr\left( S^{\mathrm T}\cdot B\cdot S - \Delta S^{\mathrm T}_-\cdot B\cdot S + \Delta S^{\mathrm T}_+\cdot B\cdot S - S^{\mathrm T}\cdot B\cdot\Delta S_- \right.\\
& \quad\qquad\quad + S^{\mathrm T}\cdot B\cdot\Delta S_+ +\Delta S^{\mathrm T}_-\cdot B\cdot\Delta S_- -\Delta S^{\mathrm T}_-\cdot B\cdot\Delta S_+ \\
& \quad\qquad\quad \left. -\Delta S^{\mathrm T}_+\cdot B\cdot\Delta S_- + \Delta S^{\mathrm T}_+\cdot B\cdot\Delta S_+\right)\\
& = q + \frac{1}{2m}\left(-\sum_{ik}1_{\alpha \beta}^{\mathrm T} B_{ij}S_{jk} \delta_{\beta k}+\sum_{ik}1_{\alpha \beta'}^{\mathrm T} B_{i j}S_{jk}\delta_{\beta' k}\right.\\
& \quad -\sum_{jk}S_{ik}^{\mathrm T} B_{i j}1_{\alpha \beta}\delta_{\beta k}+\sum_{jk}S_{ik}^{\mathrm T} B_{i j}1_{\alpha \beta'}\delta_{\beta' k}+\sum_{ij} 1_{\alpha \beta}^{\mathrm T} B_{i j}1_{\alpha\beta}\delta_{\beta \beta}\\
& \quad \left.-\sum_{ij} 1_{\alpha \beta}^{\mathrm T} B_{i j}1_{\alpha\beta'}\delta_{\beta \beta'}-\sum_{ij} 1_{\alpha \beta'}^{\mathrm T} B_{i j}1_{\alpha\beta}\delta_{\beta \beta'}+\sum_{ij} 1_{\alpha \beta'}^{\mathrm T} B_{i j}1_{\alpha\beta'}\delta_{\beta' \beta'}\right)\\
& = q + \frac{1}{2m}\left(-1_{\beta \alpha} B_{\alpha j}S_{j\beta}+1_{\beta' \alpha} B_{\alpha j}S_{j\beta'}- S_{\beta i} B_{i \alpha}1_{\alpha \beta} \right.\\
& \quad \left. + S_{\beta' i} B_{i \alpha}1_{\alpha \beta'}+ 2 B_{\alpha \alpha}\right)\\
& = q - \frac{1}{m} B_{\alpha i}S_{i\beta} + \frac{1}{m} B_{\alpha i}S_{i\beta'}+\frac{1}{m} B_{\alpha \alpha}\:,
\end{eqnarray*}
where we have exploited the fact that $B$ is symmetric.

Next, define the $N \times r$ matrix $W$ as $W\equiv B\cdot S$,
So that its components are
\begin{equation*}
W_{ik} = B_{ij}S_{jk}\:.
\end{equation*}
Then, we have
\begin{equation*}
q' = q - \frac{1}{m} W_{\alpha \beta} + \frac{1}{m} W_{\alpha \beta'}+\frac{1}{m} B_{\alpha \alpha}\:.
\end{equation*}

Now, consider making a second change in a component of $S$. Let's indicate
with $\alpha^{(1)}$ the first node moved, which switched from community $\beta^{(1)}$
to $\beta'^{(1)}$. Then, the second change moves node $\alpha^{(2)}$ from
community $\beta^{(2)}$ to $\beta'^{(2)}$. Note that $\alpha^{(2)} \neq \alpha^{(1)}$
and $\beta'^{(2)} \neq \beta^{(2)}$.
The new value of the modularity is
\begin{eqnarray*}
q^{(2)} &= q - \frac{1}{m} B_{\alpha^{(1)} i}S_{i\beta^{(1)}} + \frac{1}{m} B_{\alpha^{(1)} i}S_{i\beta'^{(1)}} - \frac{1}{m} B_{\alpha^{(2)}i}S_{i\beta^{(2)}}\\
& \quad +\frac{1}{m} B_{\alpha^{(2)} i}S_{i\beta'^{(2)}} + \frac{1}{m} B_{\alpha^{(1)} \alpha^{(1)}} + \frac{1}{m} B_{\alpha^{(2)} \alpha^{(2)}}\\
& \quad +\frac{1}{2m}\Tr (1_{\beta^{(1)} \alpha^{(1)}} B_{\alpha^{(1)} \alpha^{(2)}} 1_{\alpha^{(2)} \beta^{(2)}} + 1_{\beta^{(2)} \alpha^{(2)}} B_{\alpha^{(2)} \alpha^{(1)}} 1_{\alpha^{(1)} \beta^{(1)}} \\
& \quad\qquad -1_{\beta^{(1)} \alpha^{(1)}} B_{\alpha^{(1)} \alpha^{(2)}} 1_{\alpha^{(2)} \beta'^{(2)}} -1_{\beta^{(2)} \alpha^{(2)}} B_{\alpha^{(2)} \alpha^{(1)}} 1_{\alpha^{(1)} \beta'^{(1)}}\\
& \quad\qquad -1_{\beta'^{(1)} \alpha^{(1)}} B_{\alpha^{(1)} \alpha^{(2)}} 1_{\alpha^{(2)} \beta^{(2)}} -1_{\beta'^{(2)} \alpha^{(2)}} B_{\alpha^{(2)} \alpha^{(1)}} 1_{\alpha^{(1)} \beta^{(1)}}\\
& \quad\qquad +1_{\beta'^{(1)} \alpha^{(1)}} B_{\alpha^{(1)} \alpha^{(2)}} 1_{\alpha^{(2)} \beta'^{(2)}} +1_{\beta'^{(2)} \alpha^{(2)}} B_{\alpha^{(2)} \alpha^{(1)}} 1_{\alpha^{(1)} \beta'^{(1)}})\\
&= q-\frac{1}{m} B_{\alpha^{(1)} i}S_{i\beta^{(1)}} +\frac{1}{m} B_{\alpha^{(1)} i}S_{i\beta'^{(1)}} -\frac{1}{m} B_{\alpha^{(2)} i}S_{i\beta^{(2)}}\\
& \quad +\frac{1}{m} B_{\alpha^{(2)} i}S_{i\beta'^{(2)}} +\frac{1}{m} B_{\alpha^{(1)} \alpha^{(1)}} +\frac{1}{m} B_{\alpha^{(2)} \alpha^{(2)}}\\
& \quad +\frac{1}{m} (B_{\alpha^{(2)} \alpha^{(1)}} \delta_{\beta^{(2)} \beta^{(1)}} -B_{\alpha^{(2)} \alpha^{(1)}} \delta_{\beta^{(2)} \beta'^{(1)}}\\
& \quad\qquad -B_{\alpha^{(2)} \alpha^{(1)}} \delta_{\beta'^{(2)} \beta^{(1)}} +B_{\alpha^{(2)} \alpha^{(1)}} \delta_{\beta'^{(2)} \beta'^{(1)}})\\
&= q-\frac{1}{m} B_{\alpha^{(1)} i}S_{i\beta^{(1)}} +\frac{1}{m} B_{\alpha^{(1)} i}S_{i\beta'^{(1)}} -\frac{1}{m} B_{\alpha^{(2)} i}S_{i\beta^{(2)}}\\
& \quad +\frac{1}{m} B_{\alpha^{(2)} i}S_{i\beta'^{(2)}} +\frac{1}{m} B_{\alpha^{(1)} \alpha^{(1)}} +\frac{1}{m} B_{\alpha^{(2)} \alpha^{(2)}}\\
& \quad +\frac{1}{m} B_{\alpha^{(2)} \alpha^{(1)}} (\delta_{\beta^{(2)} \beta^{(1)}} -\delta_{\beta^{(2)} \beta'^{(1)}} -\delta_{\beta'^{(2)} \beta^{(1)}} +\delta_{\beta'^{(2)} \beta'^{(1)}})\\
&= q^{(1)} -\frac{1}{m} B_{\alpha^{(2)} i}S_{i\beta^{(2)}} +\frac{1}{m} B_{\alpha^{(2)} i}S_{i\beta'^{(2)}} +\frac{1}{m} B_{\alpha^{(2)} \alpha^{(2)}}\\
& \quad +\frac{1}{m} B_{\alpha^{(2)} \alpha^{(1)}} \left(\delta_{\beta^{(2)} \beta^{(1)}} -\delta_{\beta^{(2)} \beta'^{(1)}} -\delta_{\beta'^{(2)}\beta^{(1)}} +\delta_{\beta'^{(2)} \beta'^{(1)}}\right)\\
&= q^{(1)} -\frac{1}{m} W_{\alpha^{(2)} \beta^{(2)}} +\frac1m W_{\alpha^{(2)} \beta'^{(2)}} +\frac{1}{m} B_{\alpha^{(2)} \alpha^{(2)}}\\
& \quad +\frac{1}{m} B_{\alpha^{(2)} \alpha^{(1)}} \left(\delta_{\beta^{(2)} \beta^{(1)}} -\delta_{\beta^{(2)} \beta'^{(1)}} -\delta_{\beta'^{(2)}\beta^{(1)}} +\delta_{\beta'^{(2)} \beta'^{(1)}}\right)\\
&= q^{(1)} -\frac{1}{m} W^{(1)}_{\alpha^{(2)} \beta^{(2)}} +\frac{1}{m} W^{(1)}_{\alpha^{(2)} \beta'^{(2)}} +\frac{1}{m} B_{\alpha^{(2)} \alpha^{(2)}}\:,
\end{eqnarray*}
where
\begin{equation*}
W^{(1)}_{\alpha^{(2)} \beta^{(2)}} = W_{\alpha^{(2)} \beta^{(2)}} -B_{\alpha^{(2)} \alpha^{(1)}} \left(\delta_{\beta^{(2)} \beta^{(1)}} -\delta_{\beta^{(2)}\beta'^{(1)}}\right)\:.
\end{equation*}

Generalizing to the $(n+1)^\mathrm{th}$ change,
\begin{equation*}
\fl q^{(n+1)} = q^{(n)} -\frac{1}{m} W^{(n)}_{\alpha^{(n+1)} \beta^{(n+1)}} +\frac{1}{m} W^{(n)}_{\alpha^{(n+1)} \beta'^{(n+1)}} +\frac{1}{m} B_{\alpha^{(n+1)} \alpha^{(n+1)}}
\end{equation*}
where
\begin{equation*}
\fl W^{(n)}_{\alpha^{(n+1)} \beta^{(n+1)}} = W^{(n-1)}_{\alpha^{(n+1)} \beta^{(n+1)}} -\sum_{p=1}^{n}B_{\alpha^{(n+1)} \alpha^{(p)}} \left(\delta_{\beta^{(n+1)} \beta^{(p)}} -\delta_{\beta^{(n+1)} \beta'^{(p)}}\right)\:.
\end{equation*}
Rewrite this as
\begin{equation*}
W^{(n)}_{\alpha^{(n+1)} \beta^{(n+1)}} = W^{(n-1)}_{\alpha^{(n+1)} \beta^{(n+1)}} -\Delta W^{(n)}_{\alpha^{(n+1)} \beta^{(n+1)}}\:,
\end{equation*}
where
\begin{equation*}
\Delta W^{(n)}_{\alpha^{(n+1)} \beta^{(n+1)}} = \sum_{p=1}^{n}B_{\alpha^{(n+1)} \alpha^{(p)}}  \left(\delta_{\beta^{(n+1)} \beta^{(p)}} -\delta_{\beta^{(n+1)} \beta'^{(p)}}\right)\:.
\end{equation*}
But then it is
\begin{eqnarray*}
\Delta W^{(n)}_{\alpha^{(n+1)} \beta^{(n+1)}} &= B_{\alpha^{(n+1)} \alpha^{(n)}} \left(\delta_{\beta^{(n+1)} \beta^{(n)}} -\delta_{\beta^{(n+1)} \beta'^{(n)}}\right)\\
& \quad +\sum_{p=1}^{n-1}B_{\alpha^{(n+1)} \alpha^{(p)}} \left(\delta_{\beta^{(n+1)} \beta^{(p)}} -\delta_{\beta^{(n+1)} \beta'^{(p)}}\right)\\
&= B_{\alpha^{(n+1)} \alpha^{(n)}} \left(\delta_{\beta^{(n+1)} \beta^{(n)}} -\delta_{\beta^{(n+1)} \beta'^{(n)}}\right)\\
& \quad +\Delta W^{(n-1)}_{\alpha^{(n+1)} \beta^{(n+1)}}\:.
\end{eqnarray*}

So, the final-tuning algorithm can be implemented as follows:
\begin{enumerate}
\item Calculate $W^{(0)}_{ab} = B_{ai}S_{ib}$ for all $a$ and $b$.
\item Calculate $q^{(1)} = q + \frac{1}{m} \left( B_{a a} - W^{(0)}_{ab} + W^{(0)}_{ab'} \right)$,
where $b$ is the starting community of node $a$,
for all $a$ and $b'$, with $b' \neq b$,
and choose the one that results in the largest value of $q^{(1)}-q$.
Define that value of $a$ to be $\alpha^{(1)}$ and that value of $b'$
to be $\beta'^{(1)}$.
\item Define $\Delta W^{(0)}_{ab} = 0$ for all $a$ and $b$.
\item Set $n=1$.
\item Calculate $\Delta W^{(n)}_{ab} = \Delta W^{(n-1)}_{ab} + B_{a\alpha^{(n)}} \left(\delta_{b \beta^{(n)}}-\delta_{b \beta'^{(n)}}\right)$
for all $a$ except $\left\lbrace \alpha^{(1)}, \ldots, \alpha^{(n)}\right\rbrace$, and all $b$.
\item Calculate $W^{(n)}_{ab} = W^{(n-1)}_{ab} - \Delta W^{(n)}_{ab}$
for all $a$ except $\left\lbrace \alpha^{(1)}, \ldots, \alpha^{(n)}\right\rbrace$,
and all $b$.
\item Calculate $q^{(n+1)} = q^{(n)} + \frac{1}{m} \left( B_{aa} - W^{(n)}_{ab} + W^{(n)}_{ab'} \right)$
for all $a$ except $\{\alpha^{(1)}, \ldots, \alpha^{(n)}\}$,
and all $b'$,
and choose the pair that results in the largest value of $q^{(n+1)}-q^{(n)}$.
Define that value of $a$ to be $\alpha^{(n+1)}$, and
that $b'$ to be $\beta'^{(n+1)}$
\item If $n+1 < N$, set $n=n+1$ and go to step~(v).
\end{enumerate}

To estimate the computational complexity of the final-tuning algorithm,
consider the complexity of each step:
\begin{itemize}
\item Step~(i) is ${\cal O}(N^2)$.
\item Steps~(ii) to~(iv) are ${\cal O}(N^2)$.
\item Steps~(v) to~(vii) are ${\cal O}(N^2)$, but are repeated ${\cal O}(N^2)$ times.
\end{itemize}
Thus, the worst case computational complexity of a final-tuning
update is ${\cal O}(N^3)$.

\section{Computational complexity of the agglomeration step}\label{compaggl}
To find complexity of the agglomeration step,
first rewrite the definition of modularity as
\begin{equation*}
q = \frac{1}{2m}\left(\sum^r_{k=1 \atop k \neq x,y}\sum_{i,j \in C_k} B_{ij} + \sum_{i,j \in C_x}B_{ij} + \sum_{i,j \in C_y}B_{ij}\right)\:,
\end{equation*}
where and $C_x$ and $C_y$ are two communities
to be merged, and $r$ is the total number of
communities. Now, merge $C_x$ and $C_y$ into
a new community $C_z$. Then, the new value of
the modularity is simply
\begin{equation*}
q' = \frac{1}{2m}\left(\sum^{r'}_{k=1 \atop k \neq z}\sum_{i,j \in C_k} B_{ij} + \sum_{i,j \in C_z}B_{ij}\right)\:,
\end{equation*}
where $r'=r-1$. Now, decompose the contribution
to the modularity coming from community $C_z$ into
the contributions of its consitutent communities
$C_x$ and $C_z$:
\begin{equation*}
\sum_{i,j \in C_z}B_{ij} = \sum_{i,j \in C_x}B_{ij} + \sum_{i,j \in C_y}B_{ij} + 2\sum_{i \in C_x}\sum_{j \in C_y} B_{ij}\:.
\end{equation*}
Therefore, the change in modularity $\delta q=q'-q$ is
\begin{equation*}
\delta q = \frac{1}{m}\sum_{i \in C_x}\sum_{j \in C_y} B_{ij}\:.
\end{equation*}
Thus, we can define an $r\times r$ matrix $W$ such that its $(i,j)$
element is the change in modularity that would result from the merger
of communities $C_i$ and $C_j$.

Now, consider merging communities $C_r$ and $C_s$. If neither
community is $C_z$, then the corresponding change in modularity
is the same as it would have been in the previous step. This
means that after each merger we only need to update the rows
and columns of $W$ corresponding to the merged communities. Without
loss of generality, assume it was $x<y$. Then, it is
\begin{eqnarray*}
 W'_{xi} &= W_{xi}+W_{yi} &\quad \forall i\neq\left\lbrace x,y\right\rbrace\\
 W'_{ix} &= W_{ix}+W_{iy} &\quad \forall i\neq\left\lbrace x,y\right\rbrace\:.
\end{eqnarray*}

So, the agglomeration algorithm can be implemented as follows:
\begin{enumerate}
 \item Build the matrix $W$.
 \item Find the largest element of $W$, $W_{ij}$, with $i<j$.
 \item Move all the nodes in $C_j$ to $C_i$.
 \item Decrease the number of communities $r$ by~1.
 \item If $r>1$, update $W$ and go to step~(ii).
\end{enumerate}

To estimate the computational complexity of the agglomeration
algorithm, consider the complexity of each step:
\begin{itemize}
 \item Step~(i) is ${\cal O}(N^2)$.
 \item Steps~(ii) to~(iv) are ${\cal O}(N^2)$, but are repeated ${\cal O}(N)$ times.
\end{itemize}
Thus, the computational complexity of an agglomeration step is ${\cal O}(N^3)$.

\end{document}